\documentclass[ reprint, aps,]{revtex4-1}
\usepackage{amssymb}
\usepackage{amsmath}
\usepackage{graphicx}
\usepackage{dcolumn}
\usepackage{bm}

\begin{document}

\title{Continuous-variable quantum teleportation with non-Gaussian entangled
states generated via multiple-photon subtraction and addition}
\author{Shuai Wang$^{1,\dagger }$, Li-Li Hou$^{1}$ , Xian-Feng Chen$^{1}$,
Xue-Fen Xu$^{2}$}
\address{$^1$ School of Mathematics and Physics, Changzhou University, Changzhou 213164, People’s Republic of China
\\$^{\dagger }$ Corresponding author: wshslxy@cczu.edu.cn}

\address{$^2$ Department of Fundamental Courses, Wuxi Institute of
Technology, Wuxi 214121, People’s Republic of China}
\begin{abstract}
We theoretically analyze the Einstein-Podolsky-Rosen (EPR) correlation, the
quadrature squeezing, and the continuous-variable quantum teleportation when
considering non-Gaussian entangled states generated by applying
multiple-photon subtraction and multiple-photon addition to a two-mode
squeezed vacuum state (TMSVs). Our results indicate that in the case of the
multiple-photon-subtracted TMSVs with symmetric operations, the
corresponding EPR correlation, the two-mode squeezing degree, the sum
squeezing, and the fidelity of teleporting a coherent state or a squeezed
vacuum state can be enhanced for any squeezing parameter $r$ and these
enhancements increase with the number of subtracted photons in the
low-squeezing regime, while asymmetric multiple-photon subtractions will
generally reduce these quantities. For the multiple-photon-added TMSVs,
although it holds stronger entanglement, its EPR correlation, two-mode
squeezing, sum squeezing, and the fidelity of a coherent state are always
smaller than that of the TMSVs. Only when considering the case of
teleporting a squeezed vacuum state does the symmetric photon addition make
somewhat of an improvement in the fidelity for large-squeezing parameters.
In addition, we analytically prove that a one-mode
multiple-photon-subtracted TMSVs is equivalent to that of the one-mode
multiple-photon-added one. And one-mode multiple-photon operations will
diminish the above four quantities for any squeezing parameter $r$.

\begin{description}
\item[PACS] 42.50.Dv
\end{description}
\end{abstract}

\maketitle

\section{Introduction}

In recent years, non-Gaussian entangled states with continuous variables as
communication resources have received more attention in quantum information
and communication technologies. This is mainly because non-Gaussian states
and non-Gaussian operations are indispensable for performing some certain
continuous-variable quantum information tasks, such as quantum entanglement
distillation \cite{1,2,3,c1,4,r1,5,6,7,8,9,10,c2}, quantum error correction
\cite{12}, and universal quantum computation \cite{13}.

Photon subtraction and addition are the typical non-Gaussian operations used
to generate non-Gaussian states with highly nonclassical properties. Agarwal
and Tara \cite{14} first studied the nonclassical properties of a
photon-added coherent state, which was implemented \cite{15} via a
nondegenerate parametric amplifier with small coupling strength. The
photon-subtracted single-mode squeezed state which can be used to
conditionally produce the Schr\"{o}dinger cat state \cite{16} was
implemented by a beam splitter with high transmissivity \cite{17}. Opatrn\'{y%
} \textit{et al}. proposed that the entanglement and the fidelity of the
quantum teleportation can be enhanced by simultaneously subtracting one
photon from both modes of a two-mode squeezed vacuum state (TMSVs) \cite{1},
whereas in Ref.\cite{2}, Cochrane \textit{et al}. showed that the
entanglement and the fidelity of a coherent state with photon subtraction
are indeed increased for any nonzero initial squeezing. Considering the
transmissivity of a beam splitter and the quantum efficiency of photon
detectors, Olivares \textit{et al}. \cite{3} further proved that the
inconclusive photon subtraction is an effective method to improve the
fidelity of a coherent state when the initial squeezing is below a certain
value. Kitagawa \textit{et al}. \cite{4} provided a detailed numerical
analysis of two-mode subtraction by on-off detectors in terms of the
explicit changes in entanglement and the fidelity of a coherent state. For a
given realistic scenario with lossy transmission channels, Zhang and Loock
\cite{9} showed that in order to improve the entanglement and the fidelity,
a constraint represented by a lower bound for the beam splitter (used for
symmetric photon subtraction) must be satisfied. The more photons are
symmetrically subtracted, the higher the entanglement and the fidelity will
be. However, this improvement will disappear for large squeezing due to the
imperfections in this kind of system. Very recently, Bartley \textit{et al}.
\cite{c3} extensively investigated different strategies for enhancing
quantum entanglement via symmetric multiple-photon subtraction from a TMSVs
in a realistic experimental scenario. At present, it has been demonstrated
in experiments \cite{r1,8} that the TMSVs can be "degaussified" by photon
subtraction, resulting in a mixed non-Gaussian state whose entanglement
degree and teleportation fidelity are improved. Very recently, Kurochkin
\textit{et al}. \cite{18} also experimentally demonstrated that by applying
the photon-subtraction operator to both modes of the TMSVs, they raised the
fraction of the two-photon component in the state, resulting in an increase
of both squeezing and entanglement by about 50\%. For a review of
quantum-state engineering with photon addition and subtraction, we refer to
Refs. \cite{19,20}.

For an entangled Gaussian resource, it is known that larger squeezing leads
to larger entanglement and higher teleportation fidelity. In addition,
Adesso and Illuminati have proven that the fidelity of teleportation and the
entanglement of the shared entangled Gaussian resource are in an exact
one-to-one correspondence \cite{b1}. In experiments, the teleportation of
Gaussian states in the standard continuous-variable
Vaidman-Braunstein-Kimble (VBK) teleportation protocol \cite{r2,r3},
including the coherent state and the squeezed vacuum state, has been
reported \cite{21,22,23,24,25}. And the teleportation of a non-Gaussian
state was carried out recently \cite{26}. In the ideal and in the realistic
VBK protocol, Dell'Anno \textit{et al}. \cite{5,6,27} systematically studied
the performance of squeezed Bell states (generalized non-Gaussian entangled
states), which coincides with photon-subtracted states and photon-added
states, where subtraction and addition operations are referred to the case
of a single photon. They found that in the non-Gaussian case, the
teleportation fidelity depends not only on the entanglement, but also on the
degree of non-Gaussianity and the degree of Gaussian affinity with the
two-mode squeezed vacuum. In particular, the Gaussian affinity is crucial.
Therefore, stronger entanglement does not mean higher teleportation fidelity
when non-Gaussian states are considered as entanglement resources, although
the entanglement is indispensable in the quantum teleportation \cite%
{5,6,7,27,28}. For implementing symmetric and asymmetric multiple-photon
subtraction and addition on both modes of the TMSVs, Navarrete-Benlloch
\textit{et al}. \cite{30} demonstrated in a idealistic scenario that the
entanglement generally increases with the number of such operations. And the
multiple-photon addition typically provides a stronger entanglement
enhancement than the multiple-photon subtraction.\ As mentioned above,
stronger entanglement does not mean higher teleportation fidelity when
non-Gaussian states are considered as entanglement resources. In this paper,
we will extend the work in Refs.\cite{5,30} and theoretically investigate
the continuous-variable quantum teleportation in the standard VBK protocol
with non-Gaussian states generated\ by the symmetric and asymmetric
multiple-photon subtraction and addition. In a idealistic scenario, we show
how symmetric or asymmetric multiple-photon operations affect the EPR
correlation and the teleportation fidelity, as well as the relations among
the Einstein-Podolsky-Rosen (EPR) correlation, the two-mode squeezing and
the teleportation fidelity.

The paper is organized as follows. In Sec. II, we provide a brief review of
the multiple-photon-subtracted TMSVs (PS-TMSVs) and the
multiple-photon-added TMSVs (PA-TMSVs), as well as their entanglement
entropy. In Sec. III, we extend the work in Refs.\cite{5,30} and further
investigate the EPR correlation, the quadrature squeezing of non-Gaussian
entangled states. In Sec. IV, considering these non-Gaussian entangled
states as entangled resources, we study the teleportation fidelity of a
coherent state and a squeezed vacuum state in the standard VBK protocol. Our
results indicate that only the symmetric multiple-photon subtraction can
effectively improve the EPR correlation, the quadrature squeezing, and the
teleportation fidelity, and these improvement are obvious in the
low-initial-squeezing regime. Finally, we investigate the performance of the
PS-TMSVs for the coherent state teleportation at a fixed EPR correlation
parameter, rather than at a fixed squeezing parameter or the entanglement
entropy. Our main results are summarized in Sec.V.

\section{Multiple-photon-added and multiple-photon-subtracted two-mode
squeezed vacuum states}

In this section, we first provide a brief review of the PA-TMSVs and
PS-TMSVs, as well as their entanglement entropy. We consider a TMSVs as an
input state, which is a typical Gaussian entangled state produced in
experiments. Theoretically, the TMSVs can be obtained by adding the two-mode
squeezed operator $S_{2}\left( r\right) =\exp \left[ r\left( a^{\dagger
}b^{\dagger }-ab\right) \right] $ with squeezing parameter $r$ to a vacuum
state with modes $A$ and $B$, that is, $\left \vert r\right \rangle
=S_{2}\left( r\right) \left \vert 0,0\right \rangle =\mathrm{sech}%
r\sum_{n=0}^{\infty }\tanh ^{n}r\left \vert n,n\right \rangle $.

By performing the multiple-photon-added operation on the TMSVs, one obtains
the normalized output state as \cite{30}%
\begin{eqnarray}
\left \vert r\right \rangle _{\text{pa}} &=&C_{k,l}^{-1/2}a^{\dagger
k}b^{\dagger l}S_{2}\left( r\right) \left \vert 0,0\right \rangle   \notag \\
&=&\sum_{n=0}^{\infty }\sqrt{\frac{\left( n+k\right) !\left( n+l\right) !}{%
\left( n!\right) ^{2}C_{k,l}\cosh ^{2}r}}\tanh ^{n}r\left \vert
n+k,n+l\right \rangle ,  \label{1}
\end{eqnarray}%
where $C_{k,l}=$ Tr$\left( a^{\dagger k}b^{\dagger l}S_{2}\left( r\right)
\left \vert 0,0\right \rangle \left \langle 0,0\right \vert S_{2}\left(
-r\right) a^{k}b^{l}\right) $ is the normalization factor of the PA-TMSVs.
Different from that in Ref.\cite{30}, here for our purpose, we first
calculate the expectation value of a general product of operators $%
a^{p}a^{\dagger q}b^{h}b^{\dagger j}$ in the TMSVs. After straightforward
calculation, we have (See Appendix)%
\begin{eqnarray}
C_{p,q,h,j} &=&\text{Tr}\left( \left \vert r\right \rangle \left \langle
r\right \vert a^{p}a^{\dagger q}b^{h}b^{\dagger j}\right)   \notag \\
&=&\sum_{m}^{\min [p,h]}\frac{p!q!h!j!\cosh ^{2p+2h}r\sinh ^{j-h}2r}{%
2^{j-h}m!\left( p-m\right) !\left( h-m\right) !}  \notag \\
&&\times \frac{\tanh ^{2m}r\delta _{p+j,q+h}}{\left( j-h+m\right) !},
\label{2}
\end{eqnarray}%
where the Kronecker delta function $\delta _{p+j,q+h}$\ means that all
off-anti-diagonal elements are zero. Obviously, when $p=q=k$ and $h=j=l$, $%
C_{p,q,h,j}$ reduces to the normalization factor $C_{k,l}$. Note that the
four quantities $n$, $n+\alpha $, $n+\beta $, and $n+\alpha +\beta $ are
nonnegative integers, and the Jacobi polynomial can be written as%
\begin{eqnarray}
P_{n}^{\left( \alpha ,\beta \right) }\left( x\right)  &=&\frac{1}{2^{n}}%
\sum_{k=0}^{n}\binom{n+\alpha }{k}\binom{n+\beta }{n-k}  \notag \\
&&\times \left( x-1\right) ^{n-k}\left( x+1\right) ^{k}.  \label{3}
\end{eqnarray}%
Thus, the normalization factor $C_{k,l}$ (without loss of generality
assuming $k\geq l$) can be written as%
\begin{equation}
C_{k,l}=k!l!\cosh ^{2k}rP_{l}^{\left( 0,k-l\right) }\left( \cosh 2r\right) .
\label{4}
\end{equation}%
When only one of the modes undergoes photon addition while the other is
unchanged (for example $l=0$), Eq.(\ref{4}) reduces to
\begin{equation}
C_{k,0}=k!\cosh ^{2k}r.  \label{e1}
\end{equation}%
Then, noting that the relation $S_{2}\left( -r\right) a^{\dagger
}S_{2}\left( r\right) =a^{\dagger }\cosh r+b\sinh r$, the normalized
one-mode added TMSVs can be written as%
\begin{equation}
\frac{1}{\sqrt{k!\cosh ^{2k}r}}a^{\dagger k}S_{2}\left( \xi \right)
\left \vert 0,0\right \rangle =S_{2}\left( \xi \right) \left \vert
k,0\right \rangle ,  \label{e2}
\end{equation}%
which is a two-mode squeezed number state.

On the other hand, subtracting multiple photons from two modes of the TMSVs,
one obtains the PS-TMSVs. Theoretically, the normalized PS-TMSVs can be
written as \cite{31}
\begin{eqnarray}
\left \vert r\right \rangle _{\text{ps}}
&=&N_{k,l}^{-1/2}a^{k}b^{l}S_{2}\left( r\right) \left \vert 0,0\right \rangle
\notag \\
&=&\sum_{n=\max [k,l]}^{\infty }\sqrt{\frac{\left( n!\right) ^{2}\mathrm{sech%
}^{2}r\tanh ^{2n}r}{N_{k,l}\left( n-k\right) !\left( n-l\right) !}}%
\left \vert n-k,n-l\right \rangle ,  \label{5}
\end{eqnarray}%
where $N_{k,l}$ is the normalization factor of the PS-TMSVs. Similarly, we
first derive the expectation value of a general product of operators $%
a^{\dagger q}a^{p}b^{\dagger j}b^{h}$ in the TMSVs (Also see Appendix)%
\begin{eqnarray}
N_{p,q,h,j} &=&\text{Tr}\left( \left \vert r\right \rangle \left \langle
r\right \vert a^{\dagger q}a^{p}b^{\dagger j}b^{h}\right)   \notag \\
&=&\sum_{m}^{\min [p,h]}\frac{p!q!h!j!\sinh ^{2p+2h}r\sinh ^{j-h}2r}{%
2^{j-h}m!\left( p-m\right) !\left( h-m\right) !}  \notag \\
&&\times \frac{\coth ^{2m}r\delta _{p+j,q+h}}{\left( j-h+m\right) !}.
\label{6}
\end{eqnarray}%
Thus, in the case of $p=q=k$ and $h=j=l$, $N_{p,q,h,j}$ reduces to the
normalization factor $N_{k,l}$, i.e.,

\begin{equation}
N_{k,l}=k!l!\sinh ^{2k}rP_{l}^{(k-l,0)}\left( \cosh 2r\right) .  \label{7}
\end{equation}%
For the case $l=0$, we have
\begin{equation*}
N_{k,0}=k!\sinh ^{2k}r.
\end{equation*}%
Then, the\ normalized one-mode subtracted TMSVs is%
\begin{equation}
\frac{1}{\sqrt{k!\sinh ^{2k}r}}a^{k}S_{2}\left( \xi \right) \left \vert
00\right \rangle =S_{2}\left( \xi \right) \left \vert 0,k\right \rangle ,
\label{e3}
\end{equation}%
which is another two-mode squeezed number state. Due to the symmetry of Eqs.(%
\ref{e2}) and (\ref{e3}), we can see that adding $k$ photons to the first
mode is equivalent to subtracting them from the same mode. In addition,
adding $k$ photons to the first mode has the same effect as subtracting them
from the second mode \cite{30}. Therefore, both the one-mode
photon-subtracted TMSVs and the one-mode photon-added TMSVs have same
quantum statistical effects.

In addition, it can be seen that Eq.(\ref{2}) [or Eq.(\ref{4})] and Eq.(\ref%
{6}) [or Eq.(\ref{7})] substantially differ for the exchange of the
hyperbolic coefficients, and they are important for further studying
nonclassical properties of both the PA-TMSVs and PS-TMSVs. Particularly,
with the help of Eqs. (\ref{2}), (\ref{4}), (\ref{6}), and (\ref{7}), it is
convenient to explore some quantum optical nonclassicalities that are
characterized by the expectation values of field operators, such as
sub-Poissonian statistics, the cross correlation, anti-bunching effects \cite%
{32}, quadrature squeezing properties (including sum squeezing and
difference squeezing) \cite{33}, as well as the entanglement characterized
by some inseparability criteria \cite{34,35,36,37}.

For our purpose, let us review the von Neumann entropy of both the PA-TMSVs
and PS-TMSVs \cite{30}. For a pure state in Schmidt form, $\left \vert \psi
\right \rangle _{AB}=\sum_{n=1}c_{n}\left \vert \alpha _{n}\right \rangle
_{A}\left \vert \beta _{n}\right \rangle _{B}$ ($c_{n}\colon $real positive)
with the orthonormal states $\left \vert \alpha _{n}\right \rangle _{A}$ and $%
\left \vert \beta _{n}\right \rangle _{B}$, the quantum entanglement is
quantified by the partial von Neumann entropy of the reduced density
operator \cite{38},%
\begin{equation}
E\left( \left \vert \psi \right \rangle _{AB}\right) =-\text{Tr}\left( \rho
_{A}\ln \rho _{A}\right) =-\sum_{n=1}c_{n}^{2}\log _{2}c_{n}^{2},  \label{8}
\end{equation}%
where the local state is given by $\rho _{A}=$Tr$_{B}\left( \left \vert \psi
\right \rangle _{AB}\left \langle \psi \right \vert \right) $. Note that Eqs.(%
\ref{1}) and (\ref{5}) are already in Schmidt form, and thus the
entanglement of the PA-TMSVs and the PS-TMSVs, respectively, can be directly
obtained \cite{30},
\begin{eqnarray}
E_{\text{pa}}^{k,l} &=&-\sum_{n=0}^{\infty }\frac{\left( n+k\right) !\left(
n+l\right) !}{\left( n!\right) ^{2}C_{k,l}\cosh ^{2}r}\tanh ^{2n}r  \notag \\
&&\times \log _{2}\frac{\left( n+k\right) !\left( n+l\right) !}{\left(
n!\right) ^{2}C_{k,l}\cosh ^{2}r}\tanh ^{2n}r,  \label{9}
\end{eqnarray}%
and
\begin{eqnarray}
E_{\text{ps}}^{k,l} &=&-\sum_{n=\max [k,l]}^{\infty }\frac{\left( n!\right)
^{2}N_{k,l}^{-1}\cosh ^{-2}r}{\left( n-k\right) !\left( n-l\right) !}\tanh
^{2n}r  \notag \\
&&\times \log _{2}\frac{\left( n!\right) ^{2}N_{k,l}^{-1}\cosh ^{-2}r}{%
\left( n-k\right) !\left( n-l\right) !}\tanh ^{2n}r.  \label{10}
\end{eqnarray}%
The amount of the entanglement of the TMSVs (in the case of $k=l=0$) is
analytically given by $E=\cosh ^{2}r\log _{2}\left( \cosh ^{2}r\right)
-\sinh ^{2}r\log _{2}\left( \sinh ^{2}r\right) $, and for other states it
can be evaluated numerically by their Schmidt coefficients. When $k=l$
(symmetric operation), one can analytically prove that $E_{\text{pa}}=E_{%
\text{ps}}$. Actually, Eq. (\ref{5}) can be rewritten as
\begin{equation}
\left \vert \xi \right \rangle _{\text{ps}}=\sum_{n=0}^{\infty }\sqrt{\frac{%
\left( n+k\right) !\left( n+k\right) !}{\left( n!\right) ^{2}N_{k,k}\tanh
^{-2k}r\cosh ^{2}r}}\tanh ^{n+k}r\left \vert n,n\right \rangle .  \label{11}
\end{equation}%
From Eq. (\ref{6}), we can derive
\begin{equation}
N_{k,k}\tanh ^{-2k}r=\sum_{m=0}^{k}\frac{\left( k!\right) ^{4}\left( \cosh
r\sinh r\right) ^{2k}}{\left[ m!\left( k-m\right) !\right] ^{2}}\coth ^{2m}r.
\label{12}
\end{equation}%
If setting $k-m=m^{\prime }$, one can immediately obtain
\begin{equation}
N_{k,k}\tanh ^{-2k}r=C_{k,k}=\sum_{m=0}^{k}\frac{\left( k!\right) ^{4}\cosh
^{4k}r\tanh ^{2m}r}{\left[ m!\left( k-m\right) !\right] ^{2}}.  \label{13}
\end{equation}%
Therefore, for a symmetric operation, both the PA-TMSVs and PS-TMSVs hold
the same set of Schmidt coefficients which leads to the exact same quantum
entanglement, a result noted in Ref. \cite{5} (\thinspace $k=l=1$). For
multiple-photon added and multiple-photon subtracted operations, as pointed
out in Ref. \cite{30}, the optimal entanglement enhancement is obtained when
the same number of operations is applied to both modes as shown in Fig. 1,
where addition and subtraction give the same entanglement enhancement. And
the entanglement increases with the number of operations. For an asymmetric
operation, their numerical analysis shows that it is always better to
perform addition rather than subtraction in order to increase the
entanglement, i.e., $E_{pa}^{k,l}>E_{ps}^{k,l}$. For a detailed discussion
of the entanglement of these non-Gaussian states, please see Ref.\cite{30}.
\begin{figure}[tbp]
\centering \includegraphics[width=8cm]{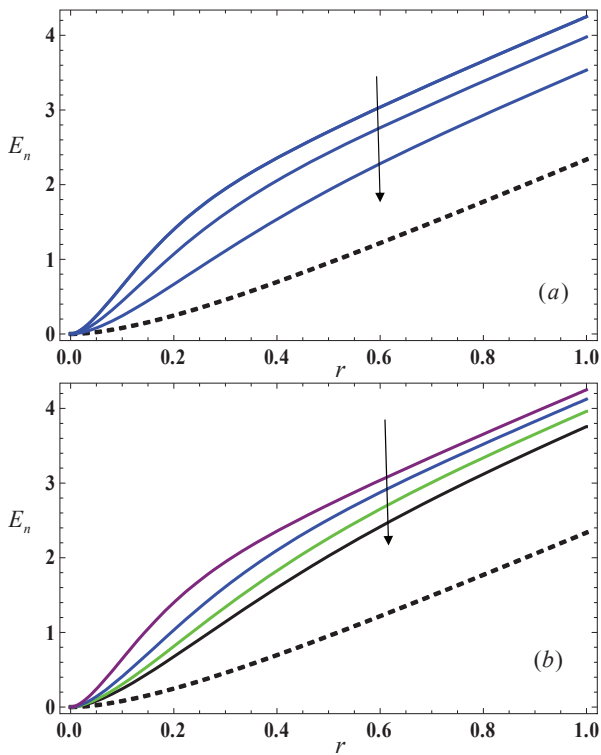}
\caption{(Color online) Entanglement entropy as a function of the squeezing
parameter $r$ for a PS-TMSVs (or PA-TMSVs) with different values of ($k,l$).
(a) PS-TMSVs or PA-TMSVs: from top to bottom, lines correspond to ($3,3$), ($%
2,2$), and ($1,1$). (b) PS-TMSVs: from top to bottom, lines correspond to ($%
3,3$), ($3,2$), ($3,1$), and ($3,0$). The black dashed curve corresponds to
the TMSVs.}
\end{figure}

\section{EPR correlation and squeezing properties}

In this section, we will further investigate the EPR correlation and the
quadrature squeezing effects (including the two-mode squeezing and the sum
squeezing) of non-Gaussian entangled states generated by multiple-photon
addition and multiple-photon subtraction.

\subsection{Einstein-Podolsky-Rosen correlation}

Besides the degree of entanglement, non-Gaussian states expressed by Eqs. (%
\ref{1}) and (\ref{5}) can be characterized by the second-order EPR
correlations between quadrature-phase components of the two modes. As
counterparts of position and momentum operators of a massive particle, the
quadrature-phase operators of each mode are defined as $X_{j}=\left(
a_{j}+a_{j}^{\dagger }\right) /\sqrt{2}$ and $P_{j}=\left(
a_{j}-a_{j}^{\dagger }\right) /(i\sqrt{2})$ ($j=A,B$). Historically,
continuous-variable entanglement originated in the paper of Einstein \textit{%
et al}., arguing on the incompleteness of quantum mechanics \cite{39}. They
proposed an ideal state which is the common eigenstate of a pair of EPR-like
operators, $X_{A}-X_{B}$ (the relative position) and $P_{A}+P_{B}$ (the
total momentum). Its explicit form is given by \cite{40,41}
\begin{equation}
\left \vert \eta \right \rangle =\exp \left[ -\frac{1}{2}\left \vert \eta
\right \vert ^{2}+\eta a^{\dagger }-\eta ^{\ast }b^{\dagger }+a^{\dagger
}b^{\dagger }\right] \left \vert 00\right \rangle ,  \label{a0}
\end{equation}%
where $\eta =\eta _{1}+i\eta _{2}$ is a complex number. In order to quantify
how well two-mode states approximate the EPR state of Eq.(\ref{a0}), one can
define the EPR correlation parameter for a generic state $\rho $ as \cite%
{42,43}
\begin{eqnarray}
\Upsilon \left( \rho \right)  &=&\Delta ^{2}\left( X_{A}-X_{B}\right)
+\Delta ^{2}\left( P_{A}+P_{B}\right)   \notag \\
&=&2\left( \left \langle a^{\dagger }a\right \rangle +\left \langle b^{\dagger
}b\right \rangle -\left \langle ab\right \rangle -\left \langle a^{\dagger
}b^{\dagger }\right \rangle +1\right)   \notag \\
&&-2\left( \left \langle a\right \rangle -\left \langle b^{\dagger
}\right \rangle \right) \left( \left \langle a^{\dagger }\right \rangle
-\left \langle b\right \rangle \right) ,  \label{14}
\end{eqnarray}%
which is the total variance of EPR-like operations $X_{A}-X_{B}$ and $%
P_{A}+P_{B}$. In the EPR state \cite{39,40,41} expressed by Eq.(\ref{a0}),
one can easily prove that the EPR correlation $\Upsilon \left( \rho \right) $
equals zero. For separable two-mode states or any classical two-mode states,
the total variance is larger than or equal to $2$. The condition
\begin{equation}
\Upsilon \left( \rho \right) \,<2,  \label{r1}
\end{equation}%
indicates quantum entanglement, which is a crucial resource for quantum
protocols using continuous variables \cite{r2,r3}.

Based on Eqs.(\ref{2}) and (\ref{6}), we can prove that $\left \langle
a\right \rangle =\left \langle a^{\dagger }\right \rangle =\left \langle
b\right \rangle =\left \langle b^{\dagger }\right \rangle =0$ and $\left \langle
\hat{a}\hat{b}\right \rangle =\left \langle \hat{a}^{\dagger }\hat{b}^{\dagger
}\right \rangle $, as well as $\left \langle a^{2}b^{2}\right \rangle
=\left \langle a^{\dagger 2}b^{\dagger 2}\right \rangle $. As a matter of
convenience, we derive the expectation values of operators $a^{\dagger
}a,b^{\dagger }b,ab,a^{2}b^{2}$ in the PS-TMSVs as follows:
\begin{eqnarray}
\left \langle a^{\dagger }a\right \rangle _{\text{ps}} &=&\frac{N_{k+1,l}}{%
N_{k,l}},\left \langle b^{\dagger }b\right \rangle _{\text{ps}}=\frac{N_{k,l+1}%
}{N_{k,l}},  \notag \\
\left \langle ab\right \rangle _{\text{ps}} &=&\frac{N_{k+1,k,l+1,l}}{N_{k,l}}%
,\left \langle a^{2}b^{2}\right \rangle _{\text{ps}}=\frac{N_{k+2,k,l+2,l}}{%
N_{k,l}},  \label{15}
\end{eqnarray}%
and
\begin{equation}
\left \langle a^{\dagger }ab^{\dagger }b\right \rangle _{\text{ps}}=\frac{%
N_{k+1,l+1}}{N_{k,l}}.  \label{16}
\end{equation}%
For the PA-TMSVs, we have%
\begin{eqnarray}
\left \langle a^{\dagger }a\right \rangle _{\text{pa}} &=&\frac{C_{k+1,l}}{%
C_{k,l}}-1,\left \langle b^{\dagger }b\right \rangle _{\text{pa}}=\frac{%
C_{k,l+1}}{C_{k,l}}-1,  \notag \\
\left \langle ab\right \rangle _{\text{pa}} &=&\frac{C_{k+1,k,l+1,l}}{C_{k,l}}%
,\left \langle a^{2}b^{2}\right \rangle _{\text{pa}}=\frac{C_{k+2,k,l+2,l}}{%
C_{k,l}},  \label{17}
\end{eqnarray}%
and
\begin{equation}
\left \langle a^{\dagger }ab^{\dagger }b\right \rangle _{\text{pa}}=\frac{%
C_{k+1,l+1}-C_{k+1,l}-C_{k,l+1}}{C_{k,l}}+1.  \label{a1}
\end{equation}%
Thus, the EPR correlation of the PS-TMSVs reads
\begin{equation}
\Upsilon \left( \rho _{\text{ps}}\right) =2\frac{%
N_{k+1,l}+N_{k,l+1}-2N_{k+1,k,l+1,l}+N_{k,l}}{N_{k,l}}.  \label{a2}
\end{equation}%
In the case of the PA-TMSVs, the EPR correlation is described in the
following form,%
\begin{equation}
\Upsilon \left( \rho _{\text{pa}}\right) =2\frac{%
C_{k+1,l}+C_{k,l+1}-2C_{k+1,k,l+1,l}-C_{k,l}}{C_{k,l}}.  \label{a3}
\end{equation}%
Particularly, when $l=0$\ (or $k=0\,$), the PS-TMSVs and the PA-TMSVs have
the same EPR correlation; then, Eqs. (\ref{a2}) and (\ref{a3}) reduce to a
simple form,%
\begin{equation}
\Upsilon \left( \rho _{\text{ps}}\right) |_{l=0}=\Upsilon \left( \rho _{%
\text{pa}}\right) |_{l=0}=\left( 2k+2\right) e^{-2r}.  \label{E1}
\end{equation}%
This is not surprising, since adding $k$\ photons to the first mode has the
same effect as subtracting them from the same mode. In the case of $k=l=0$,
Eq. (\ref{E1}) reduces to the EPR correlation of the TMSVs,
\begin{equation}
\Upsilon \left( \rho _{0}\right) =2e^{-2r},  \label{a4}
\end{equation}%
which tends to zero (the ideal EPR value) asymptotically for the squeezing
parameter $r\rightarrow \infty $. From Eq. (\ref{E1}), it clearly shows that
non-Gaussian entangled states generated by one-mode operations have a lower
EPR correlation than that of the TMSVs. Therefore, one-mode photon-addition
and photon-subtraction operations will diminish the EPR correlation.\ In the
experiment, Takahashi \textit{et al}. \cite{8} have shown that subtracting
one photon simultaneously from both modes of the TMSVs can improve the
entanglement and the EPR correlation. And subtracting a single photon from
one mode of the TMSVs enhances the entanglement while it diminishes the EPR
correlation.

In order to clearly see the EPR correlation of both the PS-TMSVs and
PA-TMSVs with different values of ($k,l$), Fig. 2 shows the symmetric
operations ($k=l$). From Fig. 2, it can be seen that the EPR correlation of
the PS-TMSVs is always larger than that of the TMSVs in the whole range of
the initial squeezing. And the EPR correlation increases with the number of
subtracted photons in the regime of low-squeezing parameter $r$. For larger
squeezing, this increase becomes negligible. However, the EPR correlation of
the PA-TMSVs is always lower than that of the input state for any values of (%
$k,l$). For asymmetric operations, both photon addition and subtraction
generally reduce the EPR correlation as shown in Fig. 3, which is different
from that of the entanglement entropy (asymmetric subtraction and addition
can also be used to improve the entanglement). In addition, our results
indicate that the EPR correlation of both the PS-TMSVs and PA-TMSVs is
optimized at symmetric operations ($k=l$). For large squeezing parameter $r$%
, the EPR correlation approaches zero, and photon addition or subtraction
cannot improve much on this. This is because when $r\rightarrow \infty $,
the EPR correlation is already closely approaching the ideal EPR value. In a
realistic scenario, the photon subtraction will indeed degrade the EPR
correlation for large initial squeezing due to the imperfections in the
systems \cite{9}.

Some works have pointed out that the larger amount of entanglement does not
always means stronger EPR correlations \cite{5,6,7,27}. And those states
holding large entanglement may not be useful to improve quantum information
processing. In the VBK protocol of quantum teleportation \cite{r2,r3}, the
quantum channel is based on the EPR correlations and the fidelity of
teleported states depends on the EPR correlations. Hence, we expect that the
quality of quantum teleportation of continuous variables can be improved by
the symmetric multiple-photon subtraction, and the fidelity can increase
with the number of subtracted photons.

\begin{figure}[tbp]
\centering \includegraphics[width=8cm]{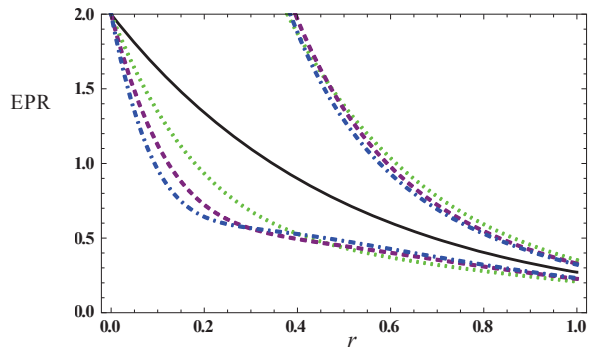}
\caption{(Color online) EPR correlation as a function of the squeezing
parameter $r$ for different non-Gaussian entangled states with $k=l$. The
three upper lines correspond to the PA-TMSVs with ($1,1$) (green dotted
line), ($2,2$) (purple dashed line), and ($3,3$) (blue dotted-dashed line).
The three lower lines correspond to the PS-TMSVs with ($1,1$) (green dotted
line), ($2,2$) (purple dashed line), and ($3,3$) (blue dotted-dashed line).
The intermediate black curve corresponds to the TMSVs.}
\end{figure}

\begin{figure}[tbp]
\centering \includegraphics[width=8cm]{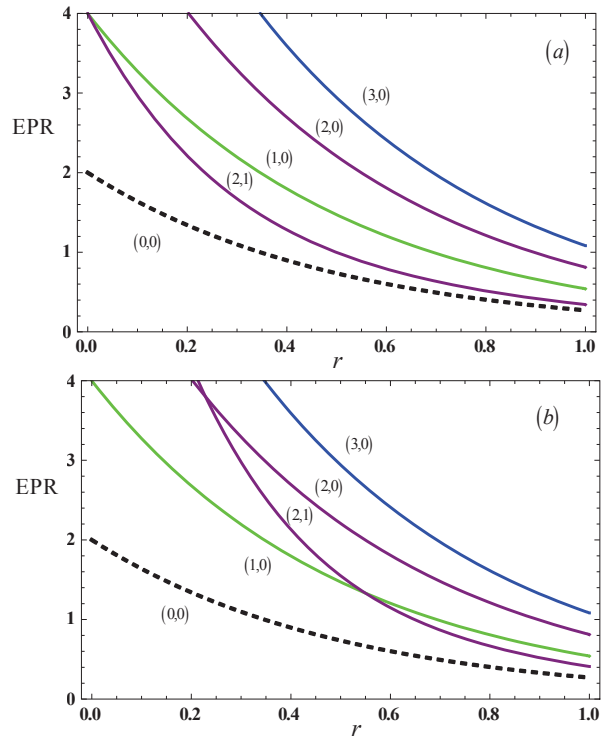}
\caption{(Color online) EPR correlation as a function of the squeezing
parameter for the states with different values of ($k,l$): ($a$) PS-TMSVs, ($%
b$) PA-TMSVs. The black dashed line corresponds to the TMSVs.}
\end{figure}

\subsection{Two-mode squeezing}

In quantum optics, squeezing is one of the earliest studied nonclassical
phenomena. There are various types of squeezing, including single-mode
squeezing, two-mode squeezing, sum squeezing and difference squeezing \cite%
{r4,r5}. Here, we study the two-mode squeezing level in which the
correlation between modes starts to play a role. For a two-mode system, the
quadrature-phase amplitudes can be expressed by $X=\left( X_{A}+X_{B}\right)
/\sqrt{2}$ and $P=\left( P_{A}+P_{B}\right) /\sqrt{2}$ ($\left[ X,P\right] =1
$), respectively. Based on Eqs.(\ref{2}) and (\ref{6}), it is easy to see
that $\left \langle a\right \rangle =\left \langle a^{\dagger }\right \rangle
=\left \langle b\right \rangle =\left \langle b^{\dagger }\right \rangle =0$ and
$\left \langle a^{2}\right \rangle =\left \langle a^{\dagger 2}\right \rangle
=\left \langle b^{2}\right \rangle =\left \langle b^{\dagger 2}\right \rangle =0$
as well as $\left \langle ab^{\dagger }\right \rangle =\left \langle a^{\dagger
}b\right \rangle =0$, which lead to $\left \langle X\right \rangle =0$ and $%
\left \langle P\right \rangle =0$. Thus, the covariances of operators $X$ and $%
P$ in the PA-TMSVs or PS-TMSVs can be written, respectively, as
\begin{equation}
\left( \Delta X\right) ^{2}=\frac{\left \langle a^{\dagger }a\right \rangle
+\left \langle b^{\dagger }b\right \rangle +2\left \langle ab\right \rangle +1}{2%
},  \label{a5}
\end{equation}%
and%
\begin{equation}
\left( \Delta P\right) ^{2}=\frac{\left \langle a^{\dagger }a\right \rangle
+\left \langle b^{\dagger }b\right \rangle -2\left \langle ab\right \rangle +1}{2%
}.  \label{a6}
\end{equation}%
According to quantum mechanics, operators $X$ and$\ P$ satisfy the
uncertainty relation $\left( \Delta X\right) ^{2}\left( \Delta P\right)
^{2}\geqslant 1/4$. When $\left( \Delta X\right) ^{2}<1/2$ or $\left( \Delta
P\right) ^{2}<1/2$, we can say there exists squeezing in the "$X$" or "$P$"
direction. Compared with Eqs.(\ref{14}), (\ref{a2}), and (\ref{a3}), we see
that the EPR correlation of non-Gaussian entangled states is four times as
much as that of the corresponding covariance of operator $P$, i.e.,
\begin{equation}
\Upsilon \left( \rho \right) =4\left( \Delta P\right) ^{2},  \label{r3}
\end{equation}%
which indicates\ that the conditions of squeezing and entanglement become
identical, which is an interesting result. Therefore, the variations of the
two-mode squeezing level for both non-Gaussian states with different values
of ($k,l$) yield the same law, as shown in Figs. 2 and 3. The subtraction
operation can enhance the degree of the two-mode squeezing, particularly in
the case of the symmetric operation. However, photon additions always weaken
the squeezing level of the TMSVs. According to Eqs. (\ref{r1}) and (\ref{E1}%
), one can see that one-mode photon-addition and photon-subtraction
operations will diminish the two-mode squeezing level and the EPR
correlation of the TMSVs.

\subsection{Sum squeezing}

Sum and difference squeezing are both higher-order, two-mode squeezing
effects \cite{33,46}. For two arbitrary modes $A$ and $B$, the sum squeezing
is associated with a so-called two-mode quadrature operator $V_{\varphi }$
of the form \cite{33}%
\begin{equation}
V_{\varphi }=\frac{1}{2}\left( e^{i\varphi }a^{\dagger }b^{\dagger
}+e^{-i\varphi }ab\right) ,  \label{a7}
\end{equation}%
where $\varphi $ is an angle made by $V_{\varphi }$ with the real axis in
the complex plane. A state is said to be sum squeezed for a $\varphi $ if
\begin{equation}
\left \langle \left( \Delta V_{\varphi }\right) ^{2}\right \rangle <\frac{1}{4}%
\left \langle a^{\dagger }a+b^{\dagger }b+1\right \rangle .  \label{a8}
\end{equation}%
From Eq. (\ref{a8}), one can define the degree of sum squeezing $S$ in the
form of normally ordered operators as follows:%
\begin{equation}
S=\frac{4\left \langle \left( \Delta V_{\varphi }\right) ^{2}\right \rangle
-\left \langle a^{\dagger }a+b^{\dagger }b+1\right \rangle }{\left \langle
a^{\dagger }a+b^{\dagger }b+1\right \rangle }.  \label{a9}
\end{equation}%
Substituting Eq.(\ref{a7}) into Eq.(\ref{a9}), we obtain $S$ as
\begin{equation}
S=\frac{2\left \langle a^{\dagger }ab^{\dagger }b\right \rangle +2\mathrm{Re}%
\left( e^{-2i\varphi }\left \langle a^{2}b^{2}\right \rangle \right) -4\mathrm{%
Re}^{2}\left( e^{-i\varphi }\left \langle ab\right \rangle \right) }{%
\left \langle a^{\dagger }a+b^{\dagger }b+1\right \rangle }.  \label{a10}
\end{equation}%
Then its negative value in the range $\left[ -1,0\right] $ indicates sum
squeezing (or higher-order nonclassicality). It is clear that $S$ has a
lower bound equal to $-1$. Hence, the closer the value of $S$ to $-1$, the
higher the degree of sum squeezing is. When $l=0$ (or $k=0\,$), the optimal
degree of the sum squeezing of both the PS-TMSVs and PA-TMSVs reduces to
that of the TMSVs at $\varphi =\pi /2$,
\begin{equation}
S_{\text{opt}}=-\frac{\left( e^{2r}-1\right) ^{2}}{\left( e^{4r}+1\right) },
\label{a12}
\end{equation}%
which is another interesting result. Thus, one-mode photon subtraction or
addition does not change the sum squeezing of the TMSVs.

The sum squeezing of the PS-TMSVs is also optimized at $\varphi =\pi /2$.
Our numerical analysis shows that the sum squeezing of the PS-TMSVs is
always larger than that of the TMSVs, and the optimal sum squeezing is
obtained for symmetric operations, as shown in Fig. 4. Thus, the degree of
the sum squeezing can be improved by the photon subtraction, particularly in
the case of symmetric operations. On the other hand, the photon addition
weakens the sum squeezing.

\begin{figure}[tbp]
\centering \includegraphics[width=8cm]{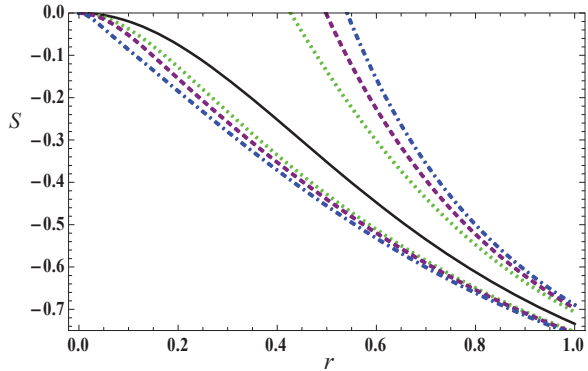}
\caption{(Color online) The sum squeezing degree $S$ as a function of the
squeezing parameter $r$ for different states. The three upper lines
correspond to the PA-TMSVs with ($1,1$) (green dotted line), ($2,2$) (purple
dashed line), and ($5,5$) (blue dot-dashed line). The three lower lines
correspond to the PS-TMSVs with ($1,1$) (green dotted line), ($2,2$) (purple
dashed line), and ($5,5$) (blue dot-dashed line). The intermediate black
curve corresponds to the TMSVs.}
\end{figure}

In this section, we demonstrate that symmetric photon subtractions can
enhance EPR correlation, the two-mode squeezing and the sum squeezing of the
TMSVs. And these quantities can be better improved with a large number of
symmetric photon subtractions in the low-initial-squeezing regime. However,
photon addition does not enhance these quantities at all, excluding the
entanglement entropy.

\section{Quantum teleportation using non-Gaussian entangled states}

Bennet \textit{et al.} \cite{47} first proposed the idea of quantum
teleportation in the discrete variable regime. After that, Vaidman \cite{r2}
put forward the idea of continuous-variable quantum teleportation. Later,
the quantum-optical protocol for the continuous-variable teleportation of
the phase-quadrature components of a light was proposed by Braunstein and
Kimble \cite{r3}. The best possible fidelity in teleporting a coherent state
without entangled resources is $1/2$ \cite{48}, so the fidelity over the
classical bound $1/2$ may be considered as a success for continuous-variable
quantum teleportation. Based on the VBK protocol, the perfect teleportation
can occur with an infinitely entangled resource that exhibits an ideal EPR
correlation [i.e., $\Delta ^{2}\left( X_{A}-X_{B}\right) +\Delta ^{2}\left(
P_{A}+P_{B}\right) \rightarrow 0$].

The success probability to teleport a pure quantum state can be described
through the teleportation fidelity, $F=$Tr$\left[ \rho _{\text{in}}\rho _{%
\text{out}}\right] $, which is a measure of how close it is between the
initial input state and the final (mixed) output quantum state. In the
formalism of the characteristic functions of the continuous variable, the
fidelity can be written as \cite{49}

\begin{equation}
F=\int \frac{d^{2}\alpha }{\pi }\chi _{\text{in}}\left( -\alpha \right) \chi
_{\text{out}}\left( \alpha \right) ,  \label{18}
\end{equation}%
where $\chi _{\text{out}}\left( \alpha \right) =\chi _{\text{in}}\left(
\alpha \right) \chi _{\text{E}}\left( \alpha ^{\ast },\alpha \right) $ \cite%
{50} is the characteristic function of the output teleported state. Here, $%
\chi _{\text{E}}\left( \alpha ^{\ast },\alpha \right) =$Tr$\left[
D_{1}\left( \alpha \right) D_{2}\left( \beta \right) \rho _{\text{E}}\right]
$ is the characteristic function of an entangled resource, and $\chi _{\text{%
in}}\left( \alpha \right) =$Tr$\left[ D\left( \alpha \right) \rho _{\text{in}%
}\right] $ is that of the input state. In the following, we consider the
non-Gaussian entangled states as entangled resources to teleport a coherent
state and a squeezed vacuum state in the standard VBK teleportation
protocol, respectively.

\subsection{Teleporting a coherent state}

Let us first consider the behavior of the fidelity for input of a coherent
state $\left \vert \beta \right \rangle $, whose characteristic function is%
\begin{equation}
\chi _{\text{coh}}\left( \alpha \right) =\exp \left[ -\frac{1}{2}\left \vert
\alpha \right \vert ^{2}+2i\mathrm{Im}\left[ \alpha \beta ^{\ast }\right] %
\right] .  \label{19}
\end{equation}%
Using the same approach as that used to derive Eq. (\ref{2}), the
characteristic functions of the PS-TMSVs and the PA-TMSVs read, respectively,%
\begin{eqnarray}
\chi _{\text{ps}}\left( \alpha ,\beta \right)  &=&\frac{\chi \left( \alpha
,\beta \right) \partial ^{2k+2l}}{N_{k,l}\partial f^{k}\partial
s^{k}\partial t^{l}\partial \tau ^{l}}e^{\left( fs+t\tau \right) \sinh
^{2}r+\left( ft+s\tau \right) \frac{\sinh 2r}{2}}  \notag \\
&&\times e^{f\left( \alpha \sinh ^{2}r-\beta ^{\ast }\frac{\sinh 2r}{2}%
\right) -s\left( \alpha ^{\ast }\sinh ^{2}r-\beta \frac{\sinh 2r}{2}\right) }
\notag \\
&&\times e^{t\left( \beta \sinh ^{2}r-\alpha ^{\ast }\frac{\sinh 2r}{2}%
\right) }  \notag \\
&&\times e^{-\tau \left( \beta ^{\ast }\sinh ^{2}r-\alpha \frac{\sinh 2r}{2}%
\right) }|_{f,s,t,\tau =0},  \label{20}
\end{eqnarray}%
and%
\begin{eqnarray}
\chi _{\text{pa}}\left( \alpha ,\beta \right)  &=&\frac{\chi \left( \alpha
,\beta \right) \partial ^{2k+2l}}{C_{k,l}\partial f^{k}\partial
s^{k}\partial t^{l}\partial \tau ^{l}}e^{\left( fs+t\tau \right) \cosh
^{2}r+\left( ft+s\tau \right) \frac{\sinh 2r}{2}}  \notag \\
&&\times e^{f\left( \alpha \cosh ^{2}r-\beta ^{\ast }\frac{\sinh 2r}{2}%
\right) -s\left( \alpha ^{\ast }\cosh ^{2}r-\beta \frac{\sinh 2r}{2}\right) }
\notag \\
&&\times e^{t\left( \beta \cosh ^{2}r-\alpha ^{\ast }\frac{\sinh 2r}{2}%
\right) }  \notag \\
&&\times e^{-\tau \left( \beta ^{\ast }\cosh ^{2}r-\alpha \frac{\sinh 2r}{2}%
\right) }|_{f,s,t,\tau =0},  \label{21}
\end{eqnarray}%
where $\chi \left( \alpha ,\beta \right) $ is the characteristic function of
the TMSVs%
\begin{equation}
\chi \left( \alpha ,\beta \right) =e^{-\frac{\cosh 2r}{2}\left( \left \vert
\alpha \right \vert ^{2}+\left \vert \beta \right \vert ^{2}\right) +\left(
\alpha \beta +\alpha ^{\ast }\beta ^{\ast }\right) \frac{\sinh 2r}{2}}.
\label{22}
\end{equation}%
It can be seen that Eqs. (\ref{20}) and (\ref{21}) substantially differ for
the exchange of the hyperbolic coefficients. Upon substituting these
characteristic functions into Eq. (\ref{18}), we can work out the fidelities
for teleporting a coherent state. For the PS-TMSVs, we have the
teleportation fidelity of a coherent state,
\begin{equation}
F_{\text{ps}}=\frac{N_{k,l}^{-1}2^{k}k!l!}{e^{-2r}+1}\left( \frac{\left(
e^{2r}-1\right) ^{2}}{4e^{2r}+4}\right) ^{l}P_{k}^{l-k,0}\left( \frac{%
e^{4r}+2e^{2r}+5}{4e^{2r}+4}\right) ,  \label{23}
\end{equation}%
where $P_{n}^{\left( \alpha ,\beta \right) }\left( x\right) $ is the Jacobi
polynomial. Different from that work of Ref.\cite{51}, here we obtain the
general expression of the fidelity for teleporting a coherent state. For the
PA-TMSVs, the teleportation fidelity of a coherent state can be written in a
simple form,
\begin{equation}
F_{\text{pa}}=\frac{\left( k+l\right) !}{4^{k+l}C_{k,l}}\frac{\left(
e^{2r}+1\right) ^{k+l}}{e^{-2r}+1},  \label{24}
\end{equation}%
which is a special case of that in Ref. \cite{52}. It can be seen that the
fidelities depend on the squeezing parameter $r$ and the number of
operations ($k,l$). Note that the fidelity is independent of the amplitudes
of the coherent state; thus Eqs. (\ref{23}) and (\ref{24}) are just the
fidelity for teleporting a vacuum state. In the case of one-mode operations
(for example $l=0$), Eqs. (\ref{23}) and (\ref{24}) reduce to
\begin{equation}
F^{\prime }=\left( \frac{1}{e^{-2r}+1}\right) ^{k+1},  \label{a13}
\end{equation}%
which is always smaller than that of the TMSVs, for any values of the
squeezing parameter $r$. Therefore, one-mode operations will also diminish
the fidelity of teleporting a coherent state.

Now, we can numerically study the behavior of the fidelity for teleporting a
coherent state by making use of the PS-TMSVs and the PA-TMSVs. In Fig. 5($a$%
), we plot the fidelity for input of a coherent state in the case of $k=l$,
i.e., the symmetric operation. From Fig. 5($a$), we can see that the
fidelity for the PS-TMSVs is always larger than that of the TMSVs, while the
fidelity for the PA-TMSVs is always smaller than that of the TMSVs, even
smaller than $1/2$ in the low-squeezing regime. For the asymmetric operation
($k\neq l$), our numerical analysis shows that both photon subtraction and
addition generally weaken the fidelity as shown in Fig. 6. Thus, the optimal
fidelity is arrived at for symmetric operations.

\begin{figure}[tbp]
\centering \includegraphics[width=8cm]{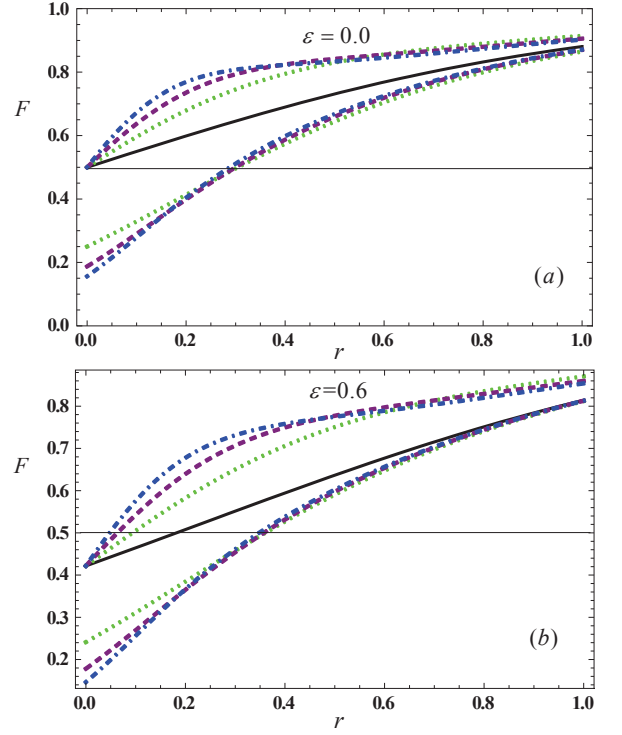}
\caption{(Color online) Fidelity as a function of the squeezing parameter
for different states The three upper lines correspond to the PS-TMSVs with ($%
1,1$) (green dotted line), ($2,2$) (purple dashed line), and ($3,3$) (blue
dotted-dashed line). The three lower lines correspond to the PA-TMSVs with ($%
1,1$) (green dotted line), ($2,2$) (purple dashed line), and ($3,3$) (blue
dotted-dashed line). The intermediate solid curve corresponds to the TMSVs. (%
$a$) The single-mode squeezing parameter $\protect \varepsilon =0.0$; ($b$) $%
\protect \varepsilon =0.6$.}
\end{figure}

\begin{figure}[tbp]
\centering \includegraphics[width=8cm]{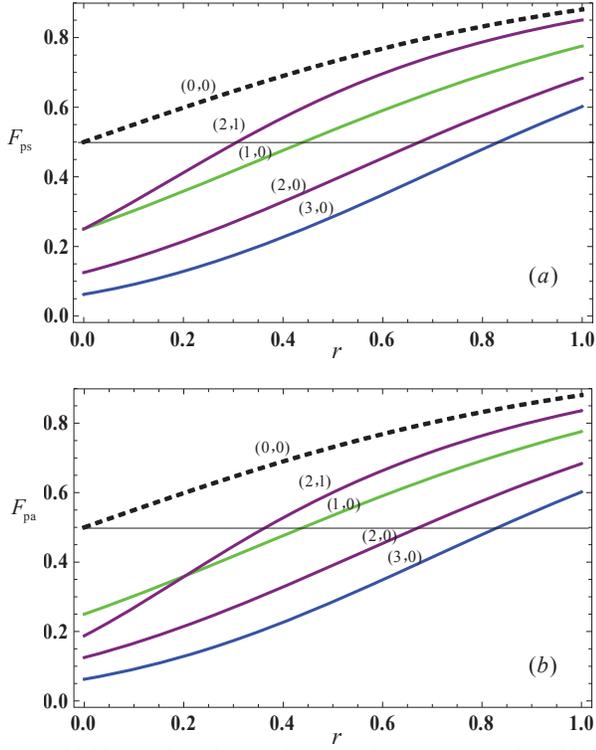}
\caption{(Color online) Fidelity as a function of the squeezing parameter
for different states: ($a$) PS-TMSVs and ($b$) PA-TMSVs. }
\end{figure}

In the VBK protocol of quantum teleportation of continuous variables, the
fidelity of teleported states depends on the EPR correlations. Thus, for
teleporting a coherent state, the higher EPR correlation means the higher
fidelity. Comparing the teleportation fidelity with the EPR correlation in
Figs. 2 and 4, one can observe that the EPR correlation and the fidelity can
be enhanced by the symmetric photon-subtraction operation in the whole
region of the squeezing parameter $r$. In addition, the teleportation
fidelity for both the PS-TMSVs and PA-TMSVs could be beyond the classical
limit of $1/2$ without the EPR correlation as shown in Figs. 2 and 4. Thus,
the teleportation fidelity, which is larger than $1/2$, does not guarantee
the existence of the EPR correlation.

\subsection{Teleporting a squeezed vacuum state}

The characteristic function of the squeezed Gaussian input state, $%
\left \vert \varepsilon \right \rangle =S\left( \varepsilon \right) \left \vert
0\right \rangle $ with the single-mode squeezing operator $S\left(
\varepsilon \right) =\exp \left[ \varepsilon \left( a^{\dagger
2}-a^{2}\right) /2\right] $ ($\varepsilon $ is the single-mode squeezing
parameter), reads%
\begin{equation}
\chi _{\text{in}}\left( \alpha \right) =\exp \left[ -\frac{\cosh
2\varepsilon }{2}\left \vert \alpha \right \vert ^{2}+\left( \alpha
^{2}+\alpha ^{\ast 2}\right) \frac{\sinh 2\varepsilon }{4}\right] .
\label{25}
\end{equation}%
With the help of the integral formula
\begin{equation}
\int \frac{d^{2}z}{\pi }e^{\zeta \left \vert z\right \vert ^{2}+\xi z+\eta
z^{\ast }+fz^{2}+gz^{\ast 2}}=\frac{1}{\sqrt{\zeta ^{2}-4fg}}e^{\frac{-\zeta
\xi \eta +\xi ^{2}g+\eta ^{2}f}{\zeta ^{2}-4fg}},  \label{26}
\end{equation}%
whose convergent condition is Re$\left( \zeta \pm f\pm g\right) <0\ $and Re$%
\left( \frac{\zeta ^{2}-4fg}{\zeta \pm f\pm g}\right) <0$, we can work out
the fidelity for teleporting a squeezed state by using the PS-TMSVs%
\begin{eqnarray}
F_{\text{ps}} &=&\frac{F_{0}}{N_{k,l}}\frac{\partial ^{2k+2l}}{\partial
f^{k}\partial s^{k}\partial t^{l}\partial \tau ^{l}}  \notag \\
&&\exp \left \{ \left( fs+t\tau \right) \sinh ^{2}r+\left( ft+s\tau \right)
\frac{\sinh 2r}{2}+\right.   \notag \\
&&\left. \frac{\left( f-\tau \right) \left( t-s\right) \left(
e^{-2r}-1\right) ^{2}\left( e^{-2r}+\cosh 2\varepsilon \right) }{4\left(
2e^{-2r}\cosh 2\varepsilon +e^{-4r}+1\right) }+\right.   \notag \\
&&\left. \frac{\left[ \left( f-\tau \right) ^{2}+\left( t-s\right) ^{2}%
\right] \left( e^{-2r}-1\right) ^{2}\sinh 2\varepsilon }{8\left(
2e^{-2r}\cosh 2\varepsilon +e^{-4r}+1\right) }\right \} |_{f,s,t,\tau =0},
\label{27}
\end{eqnarray}%
where $F_{0}$ is the fidelity for the TMSVs,%
\begin{equation}
F_{0}=\sqrt{\frac{1}{2e^{-2r}\cosh 2\varepsilon +e^{-4r}+1}}.  \label{28}
\end{equation}%
It can be seen that the fidelity is not only dependent on the squeezing
parameter $r$ and the number of subtracted photons ($k,l$), but also on the
single-mode squeezing parameter $\varepsilon $. Because of the arbitrary
order partial derivatives in Eq. (\ref{27}), finding a general expression
presents challenges.$\allowbreak $ When the squeezing parameter $\varepsilon
=0$, Eq. (\cite{27}) reduces to Eq. (\ref{23}), i.e., the fidelity of a
coherent state. For the PA-TMSVs, we obtain the fidelity of teleportation of
a squeezed vacuum state as
\begin{eqnarray}
F_{\text{pa}} &=&\frac{F_{0}}{C_{k,l}}\frac{\partial ^{2k+2l}}{\partial
f^{k}\partial s^{k}\partial t^{l}\partial \tau ^{l}}  \notag \\
&&\exp \left \{ \left( fs+t\tau \right) \cosh ^{2}r+\left( ft+s\tau \right)
\frac{\sinh 2r}{2}+\right.   \notag \\
&&\left. \frac{\left( f-\tau \right) \left( t-s\right) \left(
e^{-2r}+1\right) ^{2}\left( e^{-2r}+\cosh 2\varepsilon \right) }{4\left(
2e^{-2r}\cosh 2\varepsilon +e^{-4r}+1\right) }+\right.   \notag \\
&&\left. \frac{\left[ \left( f-\tau \right) ^{2}+\left( t-s\right) ^{2}%
\right] \left( e^{-2r}+1\right) ^{2}\sinh 2\varepsilon }{8\left(
2e^{-2r}\cosh 2\varepsilon +e^{-4r}+1\right) }\right \} |_{f,s,t,\tau =0}.
\label{29}
\end{eqnarray}%
In the case of $\varepsilon =0$, Eq.(\ref{29}) reduces to a simple form
expressed by Eq.(\ref{24}).

For the one-mode operation, Eqs.(\ref{27}) and (\ref{29}) reduce to%
\begin{equation}
F^{\prime }=\frac{F_{0}^{2k+1}}{\left( e^{-2r}\cosh 2\varepsilon +1\right)
^{-k}}\sum_{m}^{\left[ k/2\right] }\frac{k!\left( \frac{\sinh 2\varepsilon }{%
\left( \cosh 2\varepsilon +e^{2r}\right) }\right) ^{2m}}{2^{2m}\left(
m!\right) ^{2}\left( k-2m\right) !}.  \label{30}
\end{equation}%
Noting that the new expression of Legendre polynomials \cite{53} is
\begin{equation}
P_{k}\left( x\right) =x^{k}\sum_{m=0}^{[k/2]}\frac{k!\left( 1-\frac{1}{x^{2}}%
\right) ^{m}}{2^{2m}\left( m!\right) ^{2}\left( k-2m\right) !},  \label{31}
\end{equation}%
Eq.(\ref{30}) can be written as%
\begin{equation}
F^{\prime }=F_{0}^{k+1}P_{k}\left( \frac{\left( e^{-2r}\cosh 2\varepsilon
+1\right) }{\sqrt{2e^{-2r}\cosh 2\varepsilon +e^{-4r}+1}}\right) .
\label{32}
\end{equation}%
Because the factor $F_{0}^{k}P_{k}\left( x\right) $ in Eq.(\ref{32}) is
always smaller than $1$ for any values of both squeezing parameter $r$ and $%
\varepsilon $, one-mode photon subtraction and photon addition also diminish
the teleportation fidelity of a squeezed vacuum state. Obviously, when $%
\varepsilon =0$, Eq. (\ref{32}) reduces to Eq. (\ref{a13}), which is the
fidelity of teleporting a coherent state.

Compared with the coherent state, the fidelity of teleporting a squeezed
vacuum state decreases with its squeezing parameter $\varepsilon $, as shown
in Figs. 5($b$) and (7). For symmetric operation ($k=l$), the fidelity for
the PS-TMSVs is always larger than that of the TMSVs when teleporting a
squeezed vacuum state, while the fidelity for the PA-TMSVs is generally
smaller than that of the TMSVs, even smaller than 1/2 in the low-squeezing
regime. Only in the large-squeezing regime can the fidelity for the PA-TMSVs
be larger than that of the TMSVs, as shown in Fig. 5($b$). For the
asymmetric operation ($k\neq l$), both photon addition and subtraction
generally weaken the fidelity, which is similar to that behavior of
teleporting a coherent state as shown in Fig. 6. Thus, the optimal fidelity
of teleporting a squeezed vacuum state is also arrived at for symmetric
operations.

\begin{figure}[tbp]
\centering \includegraphics[width=8cm]{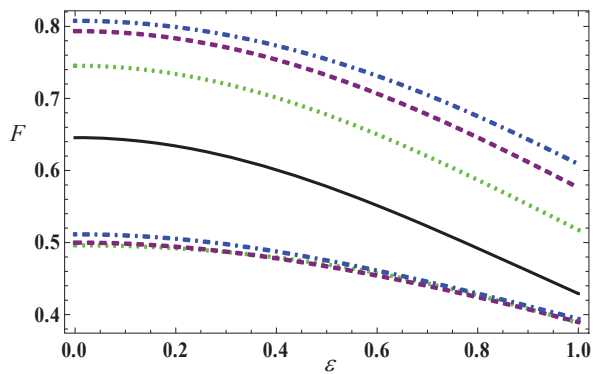}
\caption{(Color online) Fidelity as a function of the single-mode squeezing
parameter $\protect \epsilon $ with the two-mode squeezing $r=0.3$ for
different states. The three upper lines correspond to the PS-TMSVs with ($1,1
$) (green dotted line), ($2,2$) (purple dashed line), and (($3,3$) (blue
dot-dashed line). The three lower lines correspond to the PA-TMSVs with ($1,1
$) (green dotted line), $(2,2$) (purple dashed line), and ($3,3$) (blue
dot-dashed line). The intermediate black solid curve corresponds to the
TMSVs.}
\end{figure}

Although the numerical analysis shows that it is always better to perform
addition rather than subtraction in order to increase the entanglement, only
non-Gaussian entangled states generated by symmetric photon subtraction can
result in the advantage in the teleportation fidelity of a coherent or a
squeezed vacuum state, compared to just using the corresponding Gaussian
TMSVs with the same initial-squeezing parameter $r$. It has been known that
in the non-Gaussian case, the teleportation fidelity becomes a highly
complicated function of three variables: the entanglement, the degree of
non-Gaussianity, and the degree of Gaussian affinity \cite{5,6,27}. Thus the
optimal teleportation fidelity does not, in general, correspond to the
maximal entanglement. In the following, we consider the PS-TMSVs as an
entangled resource to teleport a coherent state. In Table I, we fix the EPR
correlation parameter $\Upsilon \left( \rho \right) =1.0$, and then we can
obtain the corresponding initial-squeezing parameter and the teleportation
fidelity for different ($k,l$), as well as the von Neumann entropy. From
Table I, we can see that the non-Gaussian entangled states generated by the
symmetric subtraction almost hold the same fidelity as the fixed EPR
correlation. The fidelity for the TMSVs is just a little higher than that
for the PS-TMSVs. Although the optimal entangled resource for the
teleportation of a coherent state via the ideal VBK scheme actually reduces
to the TMSVs, it is at a price, i.e., the need of the
higher-initial-squeezing parameter. In addition, Table I clearly shows again
that stronger entanglement does not mean higher teleportation fidelity even
in the multiple-photon-subtraction scheme, which is an important
illustration of the general results for the squeezed Bell states that
coincides with photon-subtracted states \cite{5}. In the limit of infinite
squeezing, since the TMSVs tends to become the ideal EPR state with perfect
EPR correlations, the fidelity of teleportation also approaches unity. Up to
the present, the largest achievable two-mode squeezing in a stable optical
configuration is about $r\approx 1.15$\ (i.e., about 10 dB) \cite{54}.
Hence, techniques that improve the performance of the teleportation without
demanding higher initial squeezing are still useful in quantum information.
In this regard, for a given two-mode squeezing parameter $r$, the TMSVs
which is engineered by non-Gaussian operation, such as symmetric
multiple-photon subtraction, contains more of the two-mode squeezing or
holds a higher EPR correlation. Thus non-Gaussian entangled states may still
be advantageous for the quantum teleportation.

\begin{table}[bph]
\begin{tabular}{|c|c|c|c|c|c|c|}
\hline
$~(k,l)~$ & ~$(2,2)$~ & ~$(1,1)$~ & ~$(0,0)$~ & ~$(1,0)$~ & ~$(2,1)$~ & ~$%
(2,0)$~ \\ \hline
$r $ & $0.1226$ & $0.1798$ & $0.3462$ & $0.6931$ & $0.5000$ & $0.8959$ \\
\hline
$F_{\text{ps}} $ & $0.6632$ & $0.6637$ & $0.6665$ & $0.6400$ & $0.6379$ & $%
0.6300$ \\ \hline
$E_{\text{ps}}^{k,l} $ & $0.5841$ & $0.5755$ & $0.5662$ & $2.094$ & $2.0925$
& $3.1624$ \\ \hline
\end{tabular}%
\caption{The fidelity varies for some different PS-TMSVs with given the EPR
correlation parameter $\Upsilon \left( \protect \rho \right) =1.0$. The
required initial-squeezing parameter $r$ and the corresponding von Neumann
entropies are also presented.}
\end{table}

On the other hand, it may be interesting to investigate the performance of
different non-Gaussian entangled states for teleporting a Gaussian state at
the fixed entanglement entropy, rather than at the fixed squeezing
parameter. When making the comparison at the fixed entanglement entropy,
Kogias \textit{et al}. \cite{55} found in all considered cases that, within
the general squeezed Bell-like class, the optimal resource state for
teleportation of input ensembles of Gaussian states via the gain-optimized
VBK scheme actually does always reduce to the TMSVs. In Fig. 8, we draw the
teleportation fidelity of coherent states as a function of the EPR
correlation and the von Neumann entropy, respectively. Figure 8($a$) shows
that if the EPR correlation parameter $\Upsilon \left( \rho \right) $\ is
smaller than a threshold value (about $0.8$), the optimal entangled resource
for the teleportation of a coherent state via the ideal VBK scheme reduces
to the PS-TMSVs generated by symmetric operation.\ When $\Upsilon \left(
\rho \right) >0.8$, the optimal entangled resource reduces to the TMSVs, for
example, $\Upsilon \left( \rho \right) =1.0$\ in table I. On the contrary,
at fixed the von Neumann entropy, the optimal entangled resource does always
reduce to the TMSVs, as shown in Fig. 8($b$), which is consistent with that
in Ref. \cite{55}. From those results in Refs. \cite{5,6,27,55} and our
results in the present work, it is clearly seen that such conclusion is
strongly dependent on the terms of comparison. This is mainly because in the
non-Gaussian case the teleportation fidelity depends not only on the
entanglement, but also on the degree of non-Gaussianity and the degree of
Gaussian affinity \cite{5,6,27}. For an entangled Gaussian resource, the
teleporation fidelity depends only on the entanglement, and both quantities
are in an exact one-to-one correspondence \cite{b1}.
\begin{figure}[tbp]
\centering \includegraphics[width=8cm]{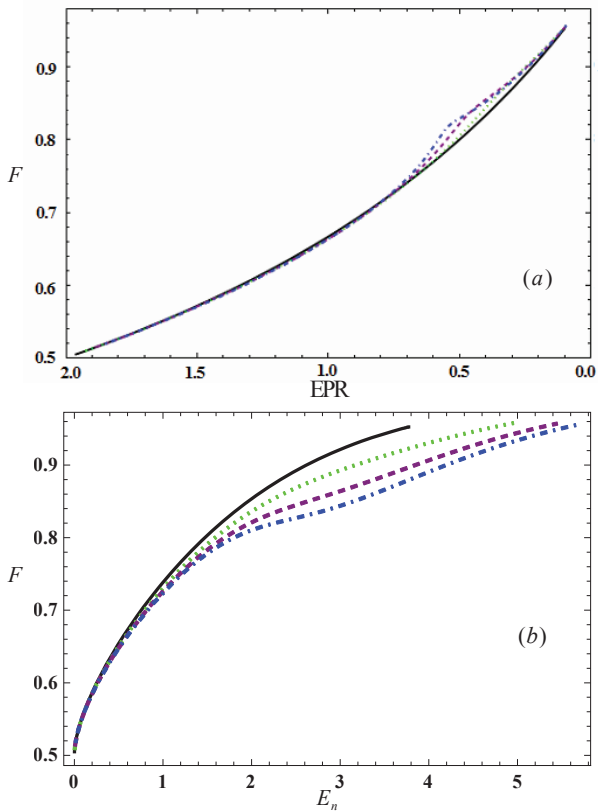}
\caption{(Color online) Fidelity of coherent teleportation with the PS-TMSVs
entangled resource. ($a$) At fixed EPR parameter; ($b$) At the fixed
entanglement entropy. These lines correspond to the PS-TMSVs with (($0,0$):
black line); (($1,1$): green dotted line), (($2,2$): purple dashed line) and
(($3,3$): blue dotted-dashed line).}
\end{figure}

\section{Conclusions}

In summary, we have shown that the symmetric multiple-photon subtraction $%
\left( k=l\right) $ can enhance the EPR correlation, the quadrature
squeezing and the teleportation fidelity of a two-mode squeezed vacuum state
(TMSVs) in the whole region of the initial-squeezing parameter $r$. Those
enhancements are more distinct in the low-initial-squeezing regime, and
increase with the number of subtracted photons. The asymmetric operation
generally diminishes the EPR correlation, the two-mode squeezing, the sum
squeezing and the teleportation fidelity, although it can enhance the
entanglement. For any values of ($k,l$), the multiple-photon addition can
better increase the degree of entanglement while it diminishes the EPR
correlation, the two-mode squeezing, the sum squeezing, and the fidelity for
teleporting a coherent state at the same time. Thus, in the
multiple-photon-subtraction or multiple-photon-addition schemes, our results
clearly show again that the entanglement enhancement does not imply that the
teleportation fidelity must be improved. The reason is that the improvement
of the fidelity is due to a balancing of three different features: the
entanglement content of the resources, their amount of non-Gaussianity, and
the degree of Gaussian affinity \cite{5}. When considering the case of
teleporting a squeezed vacuum state $\left \vert \varepsilon \right \rangle $,
the symmetric photon addition makes somewhat of an improvement on the
fidelity for large squeezing parameters $r$ and $\varepsilon $. For both the
PS-TMSVs and PA-TMSVs, the four quantities, including the optimal
entanglement, the optimal EPR correlation, the optimal quadrature squeezing,
and the optimal teleportation fidelity, always prefer symmetrical
arrangements of photon addition or subtraction on the two modes. Our results
indicate that the symmetric multiple-photon subtraction may be more useful
than the photon addition in continuous-variable quantum information
processing.

At present, the best experimentally realized non-Gaussian entangled resource
for continuous-variable teleportation is the photon-subtracted squeezed
states \cite{r1,17}. In a realistic photon-subtraction scenario, the finite
transmission coefficient of the beam splitter and the losses in the bosonic
channels, as well as the imperfection in photon-detection techniques, have a
degrading effect on the output entanglement and the fidelity of the coherent
teleportation \cite{3,4,8,9}. Due to these imperfections in these kinds of
systems, only when the initial squeezing is below a certain value can the
symmetric subtraction be used to improve the entanglement, the EPR
correlation and the fidelity of the coherent teleportation. In the
large-initial-squeezing regime, those imperfections in the
photon-subtraction scheme can make idealistic models qualitatively wrong:
for example, entanglement may decrease instead of increasing, and EPR
correlations may degrade instead of improving (as shown in Refs.\cite%
{3,4,8,9}). On the other hand, with the development of the techniques of
quantum-state engineering, it can be possible to minimize those
imperfections, particularly those imperfections in the lossy transmission
channels and the photon-detection techniques. Recently, Dell'Anno \textit{et
al} \cite{56} introduced and discussed a novel set of tunable non-Gaussian
entangled resources which contains the theoretical squeezed Bell state, as
well as an efficient scheme for their experimental generation. They find
that optimized tunable non-Gaussian resources can continue to outperform the
corresponding Gaussian resources in the realistic scenario, and even extend
to the large-initial-squeezing regime. Therefore, our theoretical results
derived in terms of the idealistic multiple-photon-subtraction and
multiple-photon-addition schemes are still meaningful.

In addition, we have analytically proved that the one-mode
multiple-photon-subtracted TMSVs is equivalent to that of the one-mode
multiple-photon-added one. For the one-mode operation, we have derived
analytical expressions of the EPR correlation, two-mode squeezing, sum
squeezing, and the teleportation fidelity, respectively. These analytical
expressions clearly represent that one-mode multiple-photon operations do
not enhance them at all, and even diminish them. Finally, we have proved
that the EPR correlation of the PA-TMSVs and PS-TMSVs is four times as much
as that of the corresponding quantum fluctuation $\left( \Delta P\right) ^{2}
$, which indicates that the conditions of the two-mode squeezing and the
entanglement become identical, which is an interesting result.

\section*{Acknowledgments}

This work is supported by the National Natural Science Foundation of China
(Grant Nos.11404040 and 11174114), and the Natural Science Foundation of
Jiangsu Province of China (Grant No. BK20140253).

\section*{Appendix: Derivation of Eqs.(\protect \ref{2}) and (\protect \ref{6})%
}

\bigskip In the Fock space, the TMSVs can be written as%
\begin{equation}
S_{2}\left( \xi \right) \left \vert 00\right \rangle =\frac{1}{\cosh r}\exp %
\left[ a^{\dagger }b^{\dagger }\tanh r\right] \left \vert 00\right \rangle .
\tag{A1}
\end{equation}%
The expectation value of a general product of operators $a^{p}a^{\dagger
q}b^{h}b^{\dagger j}$ in the TMSVs reads%
\begin{equation}
C_{p,q,h,j}=Tr\left( a^{\dagger q}b^{\dagger j}S_{2}\left( \xi \right)
\left \vert 00\right \rangle \left \langle 00\right \vert S_{2}^{\dagger }\left(
\xi \right) a^{p}b^{h}\right) .  \tag{A2}
\end{equation}%
Substituting Eq.(A1) into (A2) and inserting the completeness relation of
the two-mode coherent state, as well as using the integral formulas%
\begin{equation}
\int \frac{d^{2}z}{\pi }\exp [\zeta \left \vert z\right \vert ^{2}+\xi z+\eta
z^{\ast }]=-\frac{1}{\zeta }\exp \left[ -\frac{\xi \eta }{\zeta }\right] ,
\tag{A3}
\end{equation}%
whose convergent condition is Re$\left( \zeta \right) <0$, after doing
straightforward calculation, we obtain%
\begin{align}
C_{p,q,h,j}& = & & \frac{1}{\cosh ^{2}r}\int \frac{d^{2}z_{1}d^{2}z_{2}}{\pi
^{2}}z_{1}^{p}z_{2}^{h}z_{1}^{\ast q}z_{2}^{\ast j}  \notag \\
& & & \exp \left[ \tanh r\left( z_{1}^{\ast }z_{2}^{\ast }+z_{1}z_{2}\right)
-\left( \left \vert z_{1}\right \vert ^{2}+\left \vert z_{2}\right \vert
^{2}\right) \right]   \notag \\
& = & & \frac{1}{\cosh ^{2}r}\int \frac{d^{2}z_{1}d^{2}z_{2}}{\pi ^{2}}\frac{%
\partial ^{p+q+h+j}}{\partial f^{p}\partial s^{q}\partial t^{h}\partial \tau
^{j}}  \notag \\
& & & \exp \left[ -\left( \left \vert z_{1}\right \vert ^{2}+\left \vert
z_{2}\right \vert ^{2}\right) +fz_{1}+sz_{1}^{\ast }\right.   \notag \\
& & & \left. +tz_{2}+\tau z_{2}^{\ast }+\tanh r\left( z_{1}^{\ast
}z_{2}^{\ast }+z_{1}z_{2}\right) \right] |_{f,s,t,\tau =0}  \notag \\
& = & & \frac{\partial ^{p+q+h+j}}{\partial f^{p}\partial s^{q}\partial
t^{h}\partial \tau ^{j}}\exp \left[ \left( fs+t\tau \right) \cosh
^{2}r\right.   \notag \\
& & & \left. +\left( ft+s\tau \right) \frac{\sinh 2r}{2}\right]
|_{f,s,t,\tau =0}.  \tag{A4}
\end{align}%
By the binomial theorem, we further obtain Eq.(\ref{2})%
\begin{align}
C_{p,q,h,j}& = & & \frac{\partial ^{q+h}}{\partial s^{q}\partial t^{h}}%
\left( s\cosh ^{2}r+t\frac{\sinh 2r}{2}\right) ^{p}  \notag \\
& & & \times \left( s\frac{\sinh 2r}{2}+t\cosh ^{2}r\right) ^{j}|_{s,t=0}
\notag \\
& = & & \sum_{m}^{\min [p,h]}\frac{p!q!h!j!\left( \cosh ^{2}r\right)
^{p+h-2m}}{m!\left( p-m\right) !\left( h-m\right) !}  \notag \\
& & & \times \frac{\left( \frac{\sinh 2r}{2}\right) ^{j-h+2m}\delta
_{p+j,q+h}}{\left( j-h+m\right) !}.  \tag{A5}
\end{align}

Next, the expectation value of a general product of operators $a^{\dagger
q}a^{p}b^{\dagger j}b^{h}$ in the TMSVs reads%
\begin{equation}
N_{p,q,h,j}=Tr\left( a^{p}b^{h}S_{2}\left( \xi \right) \left \vert
00\right \rangle \left \langle 00\right \vert S_{2}^{\dagger }\left( \xi
\right) a^{\dagger q}b^{\dagger j}\right) .  \tag{A6}
\end{equation}%
By using the same approach as that to derive Eq.(A5), substituting Eq.(A1)
into (A6), and inserting the completeness relation of the two-mode coherent
state for two times, we have

\begin{align}
N_{p,q,h,j}& = & & \frac{\partial ^{p+q+h+j}}{\partial f^{p}\partial
s^{q}\partial t^{h}\partial \tau ^{j}}\exp \left[ \left( fs+t\tau \right)
\sinh ^{2}r\right.  \notag \\
& & & \left. +\left( ft+s\tau \right) \frac{\sinh 2r}{2}\right]
|_{f,s,t,\tau =0}.  \tag{A7}
\end{align}%
Similarly, we finally obtain Eq.(6).%
\begin{align}
N_{p,q,h,j}& = & & \sum_{m}^{\min [p,h]}\frac{p!q!h!j!\left( \sinh
^{2}r\right) ^{p+h-2m}}{m!\left( p-m\right) !\left( h-m\right) !}  \notag \\
& & & \times \frac{\left( \frac{\sinh 2r}{2}\right) ^{j-h+2m}\delta
_{p+j,q+h}}{\left( j-h+m\right) !}.  \tag{A8}
\end{align}


\end{document}